# Effect of different packet sizes on RED performance

Stefaan De Cnodder, Omar Elloumi, Kenny Pauwels

*Traffic and Routing Technologies project*
*Alcatel Corporate Research Center, Francis Wellesplein, 1 - 2018 Antwerp, Belgium*

*Abstract*

We consider the adaptation of random early detection (RED) as an active queue management algorithm for TCP traffic in Internet gateways where different maximum transfer units (MTUs) are used. We studied the two RED variants described in [1] and point out a weakness in both. The first variant where the drop probability is independent from the packet size discriminates connections with smaller MTUs. The second variant results in a very high Packet Loss Ratio (PLR), and as a consequence low goodput, for connections with higher MTUs. We show that fairness in terms of loss and goodput can be supplied through an appropriate setting of the RED algorithm.

# I Introduction

The random early detection (RED) algorithm is becoming a *de-facto* standard for congestion avoidance in the Internet and other packet switched networks. As a consequence of the incremental deployment of RED, several algorithms based on RED have been and are still being proposed to improve its performance.

RFC2309 [2] states that RED should be used as the default mechanism for managing queues in routers unless there are good reasons to use another mechanism. To this end, strong recommendations for testing, standardization and widespread deployment of active queue management in routers, to improve the performance of today's Internet are made.

*How does RED operate?*

In order to allow transient bursts, RED randomly drops packets based on the average queue size which is estimated as follows: $avg \leftarrow (1 - w_q) \cdot avg + w_q \cdot q$, where $w_q$ is the weight used for the exponential weighted moving average (EWMA) filter and $q$ is the instantaneous queue size. For each arriving packet if $avg$ is between a minimum and a maximum threshold then the packet is dropped with a certain probability. In [1] two variants of RED are proposed, the first one (that we denote by RED_1) does not take the packet size into account when estimating the drop probability, while the second (that we denote by RED_2) weights the drop probability by the packet size. This kind of discrimination between small and large packets is intended to avoid extra delay, incurred by retransmissions, for delay sensitive interactive traffic (e.g. Telnet) which generally consists of small packets. The following table gives the steps needed for estimating the drop probability, $p_a$, on each packet arrival for RED_1 and RED_2.

TABLE I
Drop probability estimation steps for RED_1 and RED_2.

| RED_1 | RED_2 | Step |
|---|---|---|
| $count \leftarrow count + 1$ | $count \leftarrow count + 1$ | (1) |
| $p_b \leftarrow \max_p \cdot \dfrac{avg - \min_{th}}{\max_{th} - \min_{th}}$ | $p_b \leftarrow \max_p \cdot \dfrac{avg - \min_{th}}{\max_{th} - \min_{th}}$ | (2) |
| - | $p_b \leftarrow p_b \cdot L/M$ | (3) |
| $p_a \leftarrow p_b/(1 - count \cdot p_b)$ | $p_a \leftarrow p_b/(1 - count \cdot p_b)$ | (4) |

In Table I the significance of the used parameters and variables is as follows: $p_b$ is a temporarily dropping probability, $\max_p$ is an upper bound on the temporarily packet drop probability, $\min_{th}$ and $\max_{th}$ are the two thresholds limiting the region where packets are randomly dropped, $L$ is the size of the incoming packet, $M$ is the maximum packet size and $count$ is the number of accepted packet since the last drop or since $avg$ exceeded $\min_{th}$. Note that the only difference between the two algorithms is the third step in RED_2 where the temporarily dropping probability $p_b$ is weighted by the packet size.

An attractive property for RED_1 resulting from using the $count$ variable is that the drop probability is uniformly distributed[1] [1].

## II Simulations of RED with different packet sizes

In this section we show simulation results obtained when the traffic is generated by TCP sources with different packet sizes. Our simulations are performed using 5 variants of the RED algorithm. The main differences compared to RED_1 and RED_2 is the way in which the drop probability is estimated. Table I explains the basic steps for estimating the drop probability for the three new proposed variants: RED_3, RED_4 and RED_5.

---

[1] Unpublished simulation results performed by Sally Floyd show that light-tailed distributions give better performance than heavy-tailed ones. See the note in http://www.aciri.org/floyd/REDdistributions.txt for more details.

TABLE II
Drop probability estimation steps for RED_3, RED_4 and RED_5.

| RED_3 | RED_4 | RED_5 | Step |
|---|---|---|---|
| $count \leftarrow count + 1$ | - | - | (1) |
| $p_b \leftarrow \max_p \cdot \dfrac{avg - \min_{th}}{\max_{th} - \min_{th}}$ | $p_b \leftarrow \max_p \cdot \dfrac{avg - \min_{th}}{\max_{th} - \min_{th}}$ | $p_b \leftarrow \max_p \cdot \dfrac{avg - \min_{th}}{\max_{th} - \min_{th}}$ | (2) |
| - | - | - | (3) |
| $p_a \leftarrow \dfrac{p_b \cdot L}{(1 - count \cdot p_b) \cdot M}$ | $p_a \leftarrow \dfrac{p_b \cdot L}{(1 - count \cdot p_b) \cdot M}$ | $p_a \leftarrow \dfrac{p_b}{(1 - count \cdot p_b)} \cdot \left(\dfrac{L}{M}\right)^2$ | (4) |
| - | $count \leftarrow count + \dfrac{L}{M}$ | $count \leftarrow count + \left(\dfrac{L}{M}\right)^2$ | (5) |

RED_3 is proposed as an adjustment to RED_2 in order to weight the final packet drop probability, rather than the temporary one, by the packet size. RED_4 is a small modification to RED_3 aiming at conserving a uniformly dropping function by incrementing *count* by $\dfrac{L}{M}$ and moving the update of *count* after the final drop probability calculation. In order to proof this property, let $N$ be the number of incoming packets after a packet is dropped until the next drop (including the dropped packet) and $L_i$, the length of the i<sup>th</sup> incoming packet after a drop then:

$$P[N = n] = \dfrac{p_b \cdot \dfrac{L_n}{M}}{1 - \dfrac{\sum_{i=1}^{n-1} L_i}{M} \cdot p_b} \cdot \prod_{i=1}^{n-1}\left(1 - \dfrac{\dfrac{L_i}{M} \cdot p_b}{1 - \dfrac{\sum_{j=1}^{i-1} L_j}{M} \cdot p_b}\right)$$

$$= p_b \cdot \dfrac{L_n}{M} \text{ if n verifies the following condition: } \sum_{i=1}^{n} \dfrac{L_i}{M} \leq 1/p_b,$$

and $P[N = n] = 0$ if $\sum_{i=1}^{n} \dfrac{L_i}{M} > 1/p_b$. (1)

The reason for which we proposed RED_5 comes from the TCP goodput estimation formula proposed in [3]:

$$goodput \leq \frac{MSS \cdot C}{RTT \cdot \sqrt{p}}, \qquad (2)$$

where $C$ is a constant, $MSS$ is the Maximum Segment Size, $RTT$ is the Round Trip Time and $p$ is the packet drop probability. Let $MSS_1$ and $MSS_2$ be two different MSS values corresponding to two TCP connections with the same RTT then in order to achieve fairness the following equation needs to be satisfied: $\frac{p_1}{MSS_1^2} = \frac{p_2}{MSS_2^2}$, where $p_1$ and $p_2$ are respectively the drop probability for the first and the second connection.

Note that as RED_4, RED_5 retains the property of a uniform dropping function. The proof is as in equation (1) and results in the following expression of the dropping probability:

$$P[N = n] = p_b \cdot (\frac{L_n}{M})^2 \text{ if } n \text{ verifies the following condition: } \sum_{i=1}^{n} (\frac{L_i}{M})^2 \leq 1/p_b,$$
$$\text{and } P[N = n] = 0 \text{ if } \sum_{i=1}^{n} (\frac{L_i}{M})^2 > 1/p_b. \qquad (3)$$

Our simulation model is composed of 3 groups of TCP sources/destinations sharing the same network path composed of a bottleneck link of 30 Mbits/s connecting two routers. Each group is composed of 20 TCP sources/destinations supporting selective acknowledgments where the MTU for each group is respectively 1500, 750 and 375 bytes[2]. The timeout granularity was set to 200 ms. We performed two sets of simulations with small and large propagation delay values for the bottleneck links: 15 ms and 80 ms. Simulation results reporting the PLR as well as the obtained goodput for the 5 described RED variants are depicted in Table III. A PLR for each MTU value, i.e. the number of dropped packet with a given MTU over the total number of packets having that MTU, is measured. For goodput results we plot the sum of the goodput obtained by each of the 20 TCP connections having the same MTU.

TABLE III
PLR and goodput (Mbits/s) for different MTUs and RED variants

| | | PLR (%) | | | | | Goodput (Mbits/s) | | | | |
|---|---|---|---|---|---|---|---|---|---|---|---|
| Low propagation delay | MTU | RED_1 | RED_2 | RED_3 | RED_4 | RED_5 | RED_1 | RED_2 | RED_3 | RED_4 | RED_5 |
| | small | 2.40 | 0.33 | 1.03 | 0.84 | 0.34 | 3.85 | 13.16 | 6.75 | 7.17 | 10.54 |
| | medium | 2.37 | 1.36 | 1.85 | 1.75 | 1.36 | 8.25 | 12.43 | 9.81 | 9.70 | 10.32 |
| | large | 2.41 | 13.65 | 3.57 | 3.49 | 5.22 | 16.07 | 1.29 | 11.20 | 10.85 | 6.42 |
| Sum | | | | | | | 28.17 | 26.88 | 27.76 | 27.71 | 27.27 |
| Large propagation delay | MTU | RED_1 | RED_2 | RED_3 | RED_4 | RED_5 | RED_1 | RED_2 | RED_3 | RED_4 | RED_5 |
| | small | 0.80 | 0.12 | 0.34 | 0.27 | 0.10 | 3.75 | 9.03 | 5.69 | 5.90 | 9.06 |
| | medium | 0.78 | 0.41 | 0.63 | 0.55 | 0.47 | 8.06 | 11.15 | 8.85 | 9.09 | 9.28 |
| | large | 0.74 | 2.72 | 1.07 | 1.06 | 1.73 | 15.84 | 6.84 | 12.92 | 12.47 | 8.73 |
| Sum | | | | | | | 27.65 | 27.01 | 27.46 | 27.46 | 27.08 |

---

[2] Although these MTU values may be uncommon we choose them to better explain the results of the paper. These MTU values correspond to large, medium and small MTUs in Table III.

From the simulation results we can conclude that RED_1, which drops packets without taking into account their size, results in a higher goodput for large packets. The obtained throughput is consistent with the TCP goodput estimation formula (see equation (2)): the goodput doubles when the MTU is doubled. In addition to this unfairness[3] as a function of the MTU values, it should be noted that small Telnet packets could experience too high a PLR from an interactive application requirements point of view.

For RED_2 we can clearly see that the PLR is very large for large MTU values. This leads to a considerable degradation of the TCP throughput when the propagation delay is small. Due to an increased PLR for high MTU values (close to 14% for large MTUs), the number of timeouts increases considerably and leads to important intervals of inactivity waiting for a timeout to expire. The high value of the PLR for large MTUs prevents the congestion window from reaching a sufficient value to trigger the *fast retransmit fast recovery* algorithm [4] after a packet loss. This prevents connections with a high MTU values from fairly sharing the network bottleneck. This problem is less important when the propagation delay is relatively high. The reason for this unfairness is that in RED_2 weighting the drop probability by the packet size should be done for the final drop probability rather the intermediate one. Note that, as a result for the high drop probability for large packets, the sum of the goodput decreases with RED_2 compared to the other RED variants.

We can conclude that RED_3 and RED_4 result in comparable goodput and PLRs and provide a relatively good fairness when the propagation delay is small. This fairness decreases when the propagation delay is large. The PLR for RED_3 and RED_4 doubles when the MTU doubles.

---

[3] It should be mentioned that the primary goal of RED is to monitor the queue occupancy and not to achieve a perfect fairness among TCP connections. Achieving fairness is a desired property of RED.

Finally RED_5 results in a good fairness especially when the propagation delay is large. The PLR is proportional to the square of the MTU which is an expected result. From a theoretical point of view the drop probability should be weighted by the square of the ratio of the packet size over the maximum packet size. The TCP goodput estimation formula given by equation (2) holds under the assumption that all retransmissions are made upon the receipt of three duplicate acknowledgments and not after a timeout. Hence using a small value of the timer granularity and RED_5 should improve the fairness[4] when the traffic is generated by TCP sources having different MTUs.

## III Conclusions

The main results of our simulations can be summarized as follows:

- RED_1 can result in too high a PLR for small Telnet packets,

- RED_2 could lead to a severe throughput collapse for connections with high MTU values,

- RED_3 gives good results in terms of loss differentiation and avoids low throughput (as it is the case with RED_2) for bulk transfers using large MTU values,

- RED_4 gives good results in terms of loss differentiation and fairness and results in uniformly distributed drops,

- RED_5 is, from a theoretical point of view, the best RED variant to achieve fairness for TCP-friendly traffic.

Since the traffic in the Internet is a mixture of different packet sizes we strongly recommend the use of RED_4 or RED_5 which improve the PLR differentiation and do not result in throughput degradation for connections with large MTUs.

---

[4] Perfect fairness can hardly be achieved since it is likely that different TCP connections have different RTT values. The fairness, as a function of the RTT, could be improved if the router has a means to estimate the RTT of each connection which, in practice, is hardly feasible.

## References

[1] S. Floyd and V. Jacobson, "Random Early Detection gateways for Congestion Avoidance", *IEEE/ACM Transactions on Networking,* vol.1 n.4, pp. 397-413, August 1993.

[2] B. Braden et al., "Recommendations on Queue Management and Congestion Avoidance in the Internet", RFC 2309, April 1998.

[3] M. Mathis, J. Semske, J. Mahdavi and T. Ott, "The macroscopic behavior of the TCP congestion avoidance algorithm", *Computer Communication Review*, vol. 27 n. 3, pp 67-82, July 1997.

[4] W. Stevens, "TCP Slow Start, Congestion Avoidance, Fast Retransmit, and Fast Recovery Algorithms", RFC 2001, January 1997.